# The GenUI Study: Exploring the Design of Generative UI Tools to Support UX Practitioners and Beyond

Xiang 'Anthony' Chen*
UCLA HCI Research
Los Angeles, CA, USA
xac@ucla.edu

Tiffany Knearem
Google
Boston, MA, USA
tknearem@google.com

Yang Li
Google DeepMind
Mountain View, CA, USA
liyang@google.com

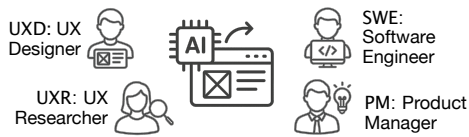

Figure 1: We conducted a formative diary study with 37 UX-related professionals who used a state-of-the-art Generative UI (GenUI) tool to conduct a week-long, individual mini-project exercise with role-specific tasks. Our findings reveal participants' workflow using the GenUI tool, how GenUI can support all and each specific roles, and existing gaps between GenUI and users' needs and expectations, which lead to design implications to inform future work on GenUI development.

## Abstract

AI can now generate high-fidelity UI mock-up screens from a high-level textual description, promising to support UX practitioners' work. However, it remains unclear how UX practitioners would adopt such Generative UI (GenUI) models in a way that is integral and beneficial to their work. To answer this question, we conducted a formative study with 37 UX-related professionals that consisted of four roles: UX designers, UX researchers, software engineers, and product managers. Using a state-of-the-art GenUI tool, each participant went through a week-long, individual mini-project exercise with role-specific tasks, keeping a daily journal of their usage and experiences with GenUI, followed by a semi-structured interview. We report findings on participants' workflow using the GenUI tool, how GenUI can support all and each specific roles, and existing gaps between GenUI and users' needs and expectations, which lead to design implications to inform future work on GenUI development.

## CCS Concepts

• **Human-centered computing** → Empirical studies in HCI.

## Keywords

GenUI, Generative AI, User Experience Design, Diary Study

## 1 Introduction

The recent advance of artificial intelligence (AI)—everything from image generation and language models—has given rise to a class of Generative UI (**GenUI**[1]) models (*e.g.*, [11, 12, 17]) that can create high-fidelity UI mock-ups based on high-level descriptions in a text prompt, promising to bring in intelligent support for the work of UX practitioners[2].

Despite its promises, the utility of GenUI models remains unclear: How would UX practitioners adopt a prompt-based approach and utilize AI-generated prototypes in their work? Can this new way of prototyping benefit other UX-related roles, such as program managers and software engineers? What are the existing challenges to realize GenUI's promises? Without answering these questions, it remains unclear whether there is value to continue advancing GenUI models; and, if these models struggle to work as expected, whether and how appropriate tool design can bridge the gap.

Despite a growing body of recent studies on the use of AI in UX design [16, 20, 24, 27, 28], to the best of our knowledge, no prior research has specifically focused on UX practitioners' use of GenUI. There remains a gap in the literature—a lack of empirical insights to inform the design of tool support that can catalyze GenUI models' utility for UX practitioners and other related roles.

To close this gap, we investigate the following research questions:

**RQ1** How would practitioners of various UX-related roles adopt GenUI in their work?

**RQ2** What are the promises and opportunities—specific aspects of their work where GenUI can provide support?

**RQ3** What are the gaps—challenges and areas for improvement—for GenUI to realize such support?

To answer these RQs, we conducted a formative study with 37 professionals within a large software company in North America

---

*This work was done when the author was a Visiting Faculty Researcher at Google DeepMind.

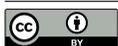

[1] There are two related terms throughout this paper: "GenUI models" and "GenUI tools"—the tool provides a user interface or IDE for accessing the model; however, at times when the distinction is less important, we simply refer to them as "GenUI".
[2] We follow [6]'s terminology and use the term "UX practitioners" to refer to both UX designers and researchers.

(Figure 1). The participants comprised of four roles typical of a software project team—11 UX designers (UXD), 7 UX researchers (UXR), 7 product managers (PM), and 12 software engineers (SWE). Participants used a state-of-the-art GenUI tool to conduct a week-long, individual mini-project exercise with role-specific tasks that provided hands-on, grounded empirical evidence. Each participant kept a daily journal of their usage and experiences with GenUI over the course of the study, followed by a semi-structured interview about how GenUI can be integral to and benefit their existing work as well as their concerns and expectations yet to be addressed.

Our main findings encompass *(i)* How participants adopted different workflows in using the GenUI tool and how they expected GenUI to be integrable into the existing workflow of their role; *(ii)* How GenUI can provide support beneficial to all roles, including early-stage ideation, saving time and effort of prototyping the first draft, and visual communication with other roles; *(iii)* How GenUI can tailor its support for non-UXD roles by democratizing UX design as well as providing additional support for tasks specific to PM, UXR, and SWE; *(iv)* Existing gaps—areas where GenUI should aim to provide better support, including problem formulation with context, assimilating users' intents, constrained generation, multimodal input & output, connecting UI elements, generated UI elements' quality, fidelity, & originality, and support for editing & iteration.

Our main contribution is a rich body of empirical findings grounded in a hands-on, project-based formative study, which surface how different roles—UX practitioners and beyond—would use GenUI in their work, the promises and opportunities for GenUI to support all and specific roles, as well as existing gaps in realizing GenUI support, leading to key implication for design to further develop GenUI models and tools.

## 2 Related Work

As we focus on GenUI—a special class of Generative AI, our literature review starts with a background of recent studies on Generative AI for UX design. Then, we briefly introduce GenUI's technical principles and a few state-of-the-art GenUI tools. Finally, GenUI is fundamentally a means of prototyping, although how much it is able to achieve towards a complete UI is still evolving as the advance of generative AI. We review previous studies of prototyping practices in industry to contextualize our understanding of how GenUI can support the work of designers and beyond.

### 2.1 Generative AI for UX Design

The recent advances in AI have sparked an increasing interest in understanding and exploring its adoption in UX design as well as existing issues and challenges [2, 13]. Stige *et al.* reviewed literature on the use of AI across the entire UX design process and summarized existing patterns of use, key issues, and open questions for future research [24]. Salminen *et al.* suggested that AI can help UX designers understand the broader context of certain user data, such as analyzing the entire think-aloud transcript to contextualize the understanding of a small excerpt [23]. Lu *et al.* [20] reviewed literature on different ways AI could support the UX design process and maps their findings on the double diamond model[3].

More related to our focus, generative AI is an emergent class of AI models capable of creating data—text, image, and more—that resemble those found in the real world. Prior to the latest waves of language and image generation models, GenAI in the past was used in a industrial design setting, *e.g.*, generating mechanical designs based on simple sketches [29].

To understand and explore the use of generative AI for UX design, Uusitalo *et al.* interviewed 10 designers to understand how they use generative AI tools in their work, such as using language models to generate personas and image generation to create illustrations of the user's journey [28]. Takaffoli *et al.*'s work [27] reported three main findings from an interview study with UX practitioners regarding their use of Generative AI: *(i)* a lack of company policies in generative AI usage, *(ii)* UX teams lack team-wide generative AI practices, and *(iii)* UX practitioners call for better training on effectively use of generative AI. The second finding is particularly relevant to our work, as we aim to explore designs that enable team-wide, role-specific GenUI practices. Li *et al.* interviewed UX practitioners in industry to learn about their perceptions of Generative AI—specifically, what it might or might not be able to assist specific aspects of work in the UX design process [16]. Relevant to our work, their findings highlight existing GenAI's limitations and challenges in supporting cross-functional team members, such as designs' hand-off to code and communication amongst various stakeholders. Naqvi *et al.* interviewed both experienced and junior designers about GenAI tool use and one common finding—related to the motivation of our study—is the questioning of whether GenAI tools can realize their marketed usefulness in design [26]. Highly related to our work, Subramonyam *et al.* studied session of software teams collaboratively using a GenAI prototyping tool, surfacing emerging approaches and a range of new challenges of using AI in the holistic design process [25].

Some research also integrated GenAI with established design tools, *e.g.*, Figma, to support UX work. Feng *et al.* developed a Figma plug-in as a probe to study how to employing LLM can help designers understand and translate requirements to specific designs [5]. PromptInfuser is another Figma plug-in that allows designers to connect UI elements to the input and output of a language model, thus supporting exploration and simulation of AI functionalities at design time [21].

Our work differs from the above studies as we focus on GenUI—an emergent subclass of generative AI. Further, we took a project-based approach to ground empirical findings in participants' hands-on experience with a state-of-the-art GenUI tool. Finally, one unique aspect of our study is the inclusion of multiple roles to explore how GenUI can benefit UX practitioners and beyond.

### 2.2 GenUI Tools

There are three representations of UI: *(i)* pixels (*e.g.*, screenshots), *(ii)* hierarchical data structure (*e.g.*, a DOM tree of Web UI elements), and *(iii)* source code that runs and renders UI (*e.g.*, code written using the React[4] framework). Correspondingly, we can approach GenUI by generating pixels of a UI screen image (*e.g.*, using Generative Adversarial Networks (GAN) [8] or Diffusion models [9]), by generating hierarchical data structure (*e.g.*, transforming an initial set of UI elements into a refined layout [17]), or generating source code (*e.g.*, using programming-by-demonstration to implement new interactive features [22] or transformers to generate a sequence of tokens representing and rendering a UI [11]). These GenUI methods, combined with recent developments of image generation and language models, have given rise to a few commercial GenUI tools, such as Uizard[5], Galileo AI[6], Visily[7], UX Pilot[8], Luny AI[9], and Figma's First Draft[10]. All these tools can take one or more types of inputs—text prompts, sketches, or screenshots—and generate UI prototypes ranging from low-fidelity wireframes to high-fidelity mock-ups. To date, no prior work has studied the use of these GenUI tools in an actual UX design setting, which is the main focus of our study. Further, the target users of these tools have not reached broader roles than UX designers; our study explored such a possibility for GenUI to support various UX-related roles. The rationale is that GenUI supports prototyping, which has broader impacts across multiple stakeholders in an industrial setting, as we review below.

---

[3]https://www.designcouncil.org.uk/our-resources/framework-for-innovation/

[4]https://react.dev/

## 2.3 Studies of Prototyping

The use of prototypes in industry predated software and UX. Industrial designers and manufacturing companies have long used physical prototypes in the process of developing a product. Yang studied how the simplicity of prototypes, the amount of time spent and where in the design process, affect the design outcome in a mechanical design scenario [30]. The main finding is that we should keep prototypes simple: prototypes with fewer parts and fewer additions of parts over time correlate with better design outcome. Lauff *et al.* conducted a longitudinal study to understand the role of physical prototypes amongst various stakeholders in a shoe manufacturing company [15]. Their findings emphasized that prototypes are communication tools [14] for a wide range of purposes, from explaining concepts to negotiating requirements and to persuading buy-in from other stakeholders. Most importantly, a prototype is constantly being "enacted through practice" by people involved based on the changing situations in the developmental process.

In the software domain, Lichter *et al.* conducted a case study of how industrial software teams adopted prototyping practices in various projects and what were the costs and effects of prototypes in these projects' success [18]. Their findings point to three foundational attributes of prototypes, that *(i)* prototyping is part of the *evolutionary* approach of software development; *(ii)* prototyping aims at creating *early working versions* of a system or application; and *(iii)* prototyping provides a medium for *communication* among team members as well as users.

Later work echoed these three characteristics when focusing on prototypes in HCI and design. Houde and Hill introduced three specific purposes to guide prototype creation, communication, and usage: understanding the *role* the system plays in its context of usage, exploring how the system looks and how users interact with it (*i.e.*,

*look and feel*), and probing technical solutions to *implement* the system [10]. Lim *et al.* proposed an anatomy of prototypes comprised of two dimensions: *filtering* certain aspects of the system design and *manifesting* the prototype in certain medium, which enables designers to take a systemic view of prototypes and prototyping practices [19]. Different than traditional software engineering that matches prototypes with requirements, Lim *et al.* pointed out that, for designers, prototypes should stimulate their thinking, helping them "communicate the rationales of their design decisions" and "frame, refine, and discover possibilities in a design space".

Amongst all the above studies, there is a shared sentiment that prototypes—whether created by humans or AI—should serve a dichotomy of purposes. First, prototypes facilitate the iterative design and development of products, from exploring early ideas, to experimenting with implementation solutions, and to evolving an early working version to a product-ready artifact [10, 18, 19]. Second, prototypes also facilitate communication by providing "a communication basis" [18], "a means to communicate ideas to others" [30], and "a communication tool" [15] for "shared understanding" [4] across different roles and stakeholders [7]. Discussed later in §5, we built on this dichotomy to distill design implications for future GenUI development.

## 3 Method

We conducted a formative study[11] to understand how different roles in a software project use GenUI in their work. The study took place from September to November, 2024. Our approach was to simulate how different UX-related roles might individually use GenUI in a mini project exercise. We gave participants access to a state-of-the-art GenUI tool and asked them to use that tool to carry out a task assignment specific to their role. This approach was in part inspired by Feng *et al.*'s work [6] where their study prompted UX practitioners to create a low-fidelity prototype while having access to Google's Teachable Machine tool.

### 3.1 Pre-Study Expert Interviews with Human-AI Interaction Experts

How UX designers would use GenUI models for prototyping might seem quite straightforward; however it is less clear, when it comes to other roles, how GenUI can play a part in supporting their work. Therefore, to define a reasonable task assignment that is appropriate for each role while able to elicit undiscovered usage of GenUI, we first conducted individual pre-study interviews with three experts in human-AI interaction.

We recruited these experts via our company's internal people search portal. They held the job titles of research scientist, creative technologist, and senior UX designer, and frequently interacted or collaborated with non-UX roles in software project teams. While the nature of their job and skill-set varied from one another, all experts identified human-AI interaction as a common theme of their work, with experiences on this particular theme ranging from six to eight years.

During the interview, one research team member facilitated the discussion while the other took notes and asked complementary questions. We started with an introduction of the project and a brief overview of several existing tools with GenUI features, including Uizard's Autodesigner 2.0, Galileo AI, and Figma's AI Design Copilot plugin. Then, we discussed the feasibility of using a GenUI tool in four different roles' work: UX designer (UXD), UX researcher (UXR), product manager (PM), and front-end software engineers (SWE). The research team converged on a written task assignment for each role, iteratively revised over the course of three interviews.

---

[5] https://app.uizard.io/
[6] https://www.usegalileo.ai/
[7] https://www.visily.ai/
[8] https://uxpilot.ai/
[9] https://www.luny-ai.com/
[10] This feature was disabled by Figma during the time of our study (https://help.figma.com/hc/en-us/articles/23955143044247-Use-First-Draft-with-Figma-AI)

[11] The authors, experts in the pre-study interview, and all the participants were from the same company.

## 3.2 Apparatus

Based on experts' feedback and on our technical assessment, we chose one state-of-the-art GenUI tool for use in the task assignment[12]. This GenUI tool, targeted at general audience, can generate high-fidelity, interactive UI mock-ups from sketches, wireframes, screenshots, or textual descriptions. The tool employs advanced computer vision algorithms to parse and interpret rough inputs, such as hand-drawn sketches, and converts them into structured UI components, including buttons, forms, and navigation bars. Its natural language processing capabilities allow users to describe desired layouts or functionalities in text, which are then transformed into corresponding UI mock-ups. Once an initial prototype is generated, the tool provides an in-situ dialog with AI for users to edit or create additional UI elements, assets, or styles.

## 3.3 Participant Recruitment & Study Design

Our study consisted of a task-based assignment that participants completed on their own followed by a semi-structured interviews to understand the participant's thoughts and experiences of using the GenUI tool. Upon completion of the study, participants received an internal incentive (*e.g.*, choice of company merchandise) for their time. This study followed the company's policy on engaging in research with human subjects.

*3.3.1 Participant recruitment.* We primarily utilized our company's internal user research recruitment system. We also used snowball sampling to get referrals from our participants to fill in gaps in user role demographics. Participants met the following criteria: employed as a UXD, PM, SWE, or UXR, and were willing to participate in all research activities during the data collection time frame. Following these strategies, we recruited a total of 37 participants, 16 men, 20 women, and 1 non-binary, aged 21 to 54, all of whom worked for a North American information technology company. Amongst the participants, there were 11 UXD, 7 PM, 12 SWE, and 7 UXR. The variation of these participant numbers, in part, reflected the distribution of roles in the company, *e.g.*, there were more UXD and SWE than PM and UXR. Although all participants were in the same company, they came from a wide variety of teams and with various experience levels. To mitigate biases towards organizational cultures and tools, we designed the task to be independent of participants' day-to-day work and instructed them to perform the tasks in their personal time.

*3.3.2 Pre-study survey.* In a short pre-study survey, we asked participants about their prior experiences with generative AI. Our goal was to get a sense of how often, if at all, they used generative AI to assist their work and their level of comfort using AI in their day-to-day work. Almost two thirds of the participants said they used language models (*e.g.*, ChatGPT and Gemini) on a daily basis. When it comes to text-to-image generation (*e.g.*, DALL-E and Midjourney), usage was less frequent—most participants used such AI weekly or monthly. Please see Appendix §A for a detailed breakdown of the AI tool usage data across the four roles. There were 7 participants who reported prior experiences using GenUI-related tools, such as Vercel's v0, Claude Artifacts, Gemini, ChatGPT, Figma's "First Draft" feature, Uizard, Musho AI, Galileo AI, and Genius AI. These participants' GenUI experiences were mainly exploratory and none had used a GenUI tool for their work as an integral part of a project.

*3.3.3 Task-based assignment procedure.* We firstly onboarded each participant via a video conference to describe the background of this project, introduce the GenUI tool's AI features via a pre-recorded tutorial video, and walked through each role's task assignment. We described to each participant a project scenario as follows: *You are part of the project team that tries to create a mobile app to facilitate mindful micro-activities throughout the day*. Then we assigned specific tasks to complete with the GenUI tool based on each participant's professional role as follows:

UXD Your task is to create a high-fidelity prototype with five screens that onboard a user to a new feature of your design. Think about this feature by filling in the following sentence: As a <user persona>, I want to <perform a specific activity> in order to <achieve a specific goal>.

PM Your task is to come up with an innovative feature of this app and create a brief product requirement document (PRD) to communicate the core idea of this feature to designers and engineers. Your PRD includes the following sections: TL;DR, Background, Goals & non-goals, Success criteria, and System diagram.

SWE You task is to develop a Web-based, minimum viable product[13] (UI-only) of this app, demonstrating an innovative feature of your design. No designer has been assigned to this project and you will need to come up with the feature on your own.

UXR Your task is to create a user research plan to investigate what valuable features this app might provide. You will decide and choose one specific method, such as diary study, competitive analysis, survey, or market research. Your user research plan includes the following: Set-up, Participants, Recruiting criteria, Timeline, Responsibilities of parties, and Estimated cost.

We asked each participant to perform the task within one week, spending a total of at least two hours—they were free to spend however much time on any day that fit their schedule. To conform with company regulations, we asked participants not to involve any work-related content and only perform the task assignment in their personal time. We instructed participants to use the GenUI features "as creatively and as much as they could benefit from it for the task".

---

[12]Rather than having participants choose from multiple tools, providing them with a single designated tool reduced their learning time and effort, while making it possible to cross-analyze participants' data to pursue more depth in our analysis based on that one GenUI tool.

[13]https://en.wikipedia.org/wiki/Minimum_viable_product

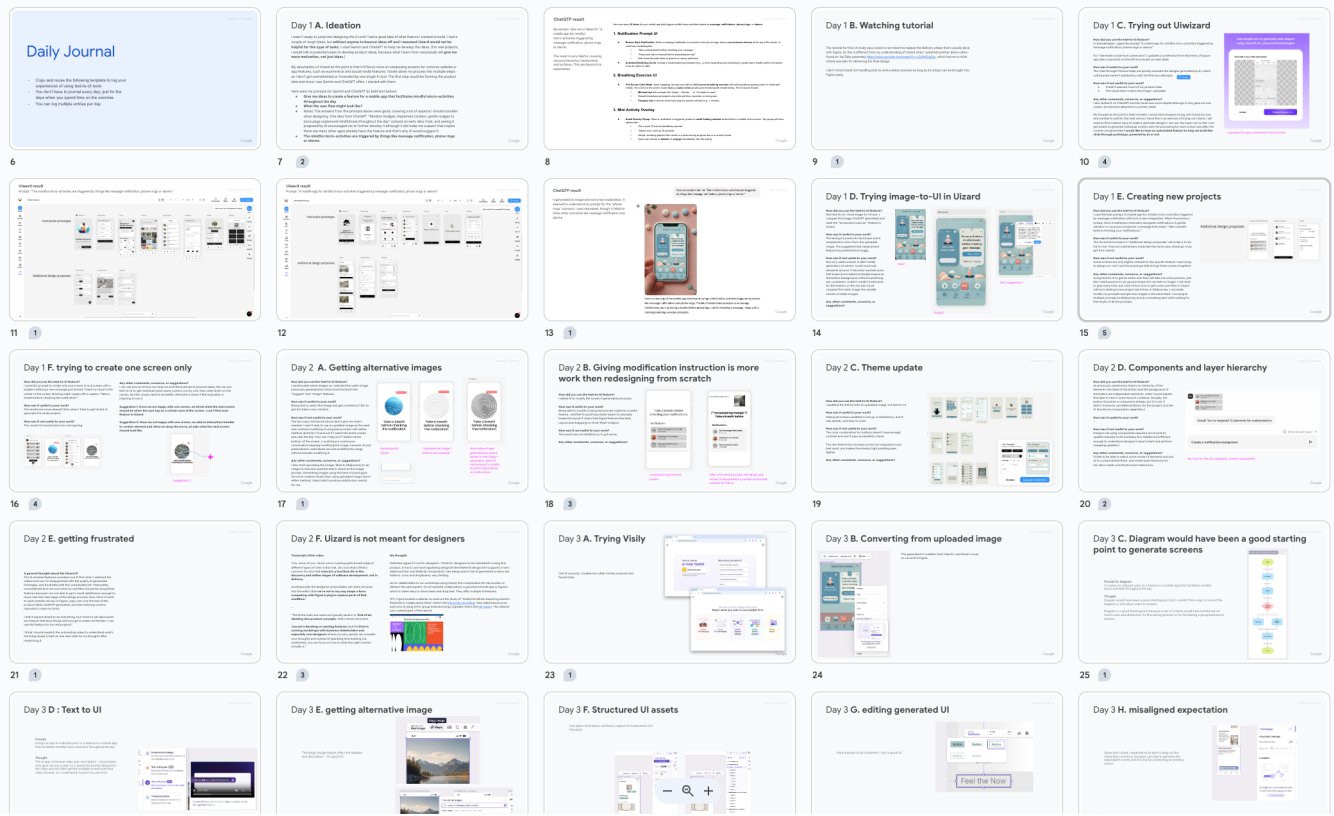

Figure 2: **UXD1's** slide deck where she kept a daily journal of her usage of, experiences with, and feedback to the GenUI tool, including screenshots of her design copied and pasted from the GenUI tool.

We provided each participant with a template (using Google Slides) for keeping a daily journal, *i.e.*, logging their activities, experiences, and feedback on the days when they worked on the task assignment, which consists of the following questions: *(i)* How did you use the GenUI feature? Include a screenshot if applicable. *(ii)* How was it useful to your work? *(iii)* How was it *not* useful to your work? *(iv)* Any other comments, concerns, or suggestions? Figure 2 shows one participant's slide deck upon her finishing the study.

*3.3.4 Semi-structured Interview.* At the end of the week-long study, we scheduled and conducted one-on-one video interviews with each participant. Each interview lasted between 30 and 45 minutes. Oral consent for participation in the interview was obtained prior to the start of each interview. We chose not to record the sessions because note-taking was sufficient for our analysis.

During the interview we first asked the participant to briefly present their work product. Next, we walked through their daily template responses, asking clarification and follow-up questions. Finally, we asked them to summarize, based on this design exercise experience, opportunities and concerns for GenUI tools to support their work as individuals or collaboratively with their team. The interviewer manually transcribed each participant's responses within the Google Slide file where the participant logged their GenUI usage and feedback. We include a detailed protocol in Appendix §C.

## 3.4 Data Analysis

We employed a grounded theory approach [1] to code the qualitative data (responses in both the daily journal and the final interview) through multiple iterations: *(i)* The first author broke the data into atomic segments and labeled each with low-level codes to identify the underlying concept(s) related to our research questions; *(ii)* The first author reviewed the initial codes and developed them into higher-level codes, which we cataloged and defined in a code book[14]; *(iii)* The second author reviewed the code book and discussed with the first author to resolve observed ambiguities, conflicts, and disagreements. By the end of these iterations, the code book contained 28 high-level codes[15]. Example codes include *workflow integration* (GenUI tools should be integrable with how users currently work and what tools they currently use in their work), *democratizing UX* (GenUI makes UX design more accessible to the broader non-UX population, allowing them to gain independence from relying on UXDs and accelerate their work), and *the last*

---
[14]We paid specific attention not to include usability-related segments that likely only existed in the very tool we used in the study, *e.g.*, the default display of the generated UI was too small.
[15]The complete code book is available in Appendix §B.

*mile problem* (Although GenUI provides a good-enough first draft, it seems to require significant effort to close the "last mile", *i.e.*, editing them to be product- or engineering-ready). Next, to further strengthen their relevance to our research questions, the first and second authors worked together to filter, sort, and converge the codes to higher-level themes to structure our main findings. Specifically, we excluded 5 codes due to their independence of GenUI (*i.e.*, their likely existence in other contexts of literature), such as *usability* and *explainability*. Next, we merged codes with close semantic proximity (*e.g.*, *support of editing* and *support of iteration & exploration*). As a result, we arrived at the following higher-level themes spanning 16 codes: how people of various roles would use GenUI in their work (§4.1); GenUI's promises and opportunities to support all roles (§4.2) as well as specific roles (§4.3); and gaps in GenUI—ongoing challenges and areas for improvement (§4.4). We report the findings that arose from these themes in the following section.

## 4 Findings
### 4.1 How Various Roles Would Use GenUI

As described earlier in §3.3.3, we assigned role-specific tasks that grounded participants' goals of using the GenUI tool. We describe two complementary sets of findings (Figure 3): *(i)* within the GenUI tool, there are two distinct workflows participants took to create UI prototypes; and *(ii)* in the broader context, how the use of GenUI can fit in users' role-specific workflow based on existing tools and practices.

> **How People Use GenUI #1: Workflow of Prototyping with GenUI**
>
> While many participants generated some screens first and then refined or add more elements or screens, others preferred creating UI incrementally one screen at a time.

Participants started their project by following the default workflow in the GenUI tool—using a high-level text prompt to generate some initial screens, then refining elements in individual screens or adding more elements or screens. However, many realized that it was not a useful starting point to generate multiple screens because generating all the elements and features at once could be too complex and overwhelming for a user who has just begun to explore the design (SWE7, PM4). Participants also found generating all screens often introduced unwanted or irrelevant features (UXD4) that misaligned with prompts— "*getting the results wrong from the get-go*" (UXD12).

In contrast, about one third of participants chose to create UI screen-by-screen through the chat interface with GenUI, which would take a text prompt and return a few design options for a screen. UXD5 used this feature to explore multiple possibilities when generating a sign-up screen. UXD4 found it helpful to describe one screen at a time as she had five specific screens in mind that she wanted to generate.

Participants also felt the needs for an iterative workflow to incrementally add details to a prompt (UXD1) in a coarse-to-grain approach (UXD4) or to revise a more clear and specific prompt based on previous generations (SWE7, SWE12). As these two needs are closely related to assimilating users' intents, we discuss them in more details in §4.4.

> **How People Use GenUI #2: Integrating GenUI with Existing Tools and Practices**
>
> GenUI can integrate itself before, after, or in between users' role-specific tasks; however, the main concern is seamlessly transferring work products between GenUI and other tools.

When asked how they would integrate the use of GenUI in their work, the most common issue participants were concerned about was integrating GenUI with other parts of their work, such as exporting results to their familiar tools, such as Figma or Google Slides (PM5, UXD2), to a study plan for user research (UXR1), to a development code base (*e.g.*, GenUI as CSS for SWE to code HTML); or, the other way round, to bring work done by Figma into the GenUI tool (UXD7). Enabling such transfer can reduce friction to adopt GenUI (SWE8), help users manage the relationships between the GenUI tool and incumbent tools like Figma (UXD12), and prevent extra work, *e.g.*, UXD using Figma to re-create what a PM designs using GenUI (PM4).

Participants did not express strong preference about whether they should use GenUI before and/or after existing tasks in their work. UXR2 wrote the research plan after using GenUI to create the flow of the app whereas PM2, PM4, and PM5 created their PRD first and then used it to describe their prompt to GenUI. PM5, in particular, expected to use GenUI to augment the PRD by adding visuals of UI mock-ups or a one-pager to summarize the main designs to other stakeholders. Others hoped GenUI can play a recurring role in their work. UXR6 would use GenUI in between iterations of user research: GenUI can help with brainstorming after conducting the initial research to give some recommendations to the team and then again supporting brainstorming before moving on to the next phase of research, *e.g.*, coming up with open questions to study. UXD2 suggested that a GenUI tool should automatically update the generated screen when there is a design update, thus maintaining consistency between the team's evolving ideas and the design files.

### 4.2 GenUI's Promises and Opportunities to Support *All Roles*

The following findings (Figure 4) encompass the expected promises and opportunities of GenUI—as an emergent class of tools to provide support that would benefit *all* roles regardless of their nature of work, including early-stage ideation, saving time and effort of prototyping, and facilitating cross-role visual communication.

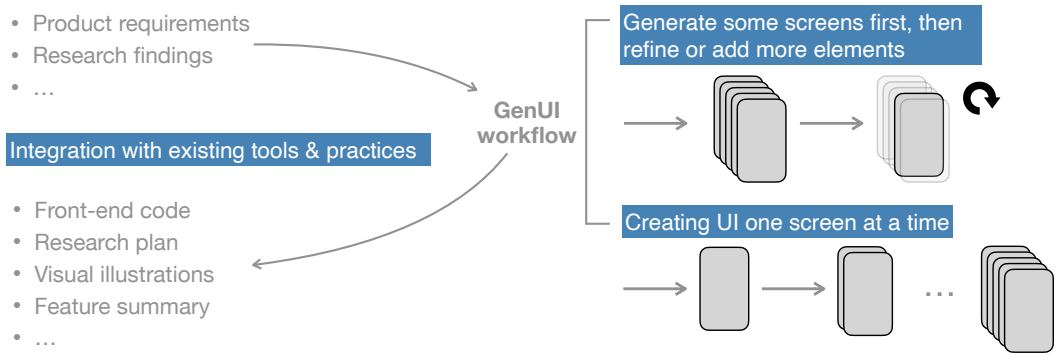

Figure 3: Our findings surface two major ways (workflow) of using the GenUI tool as well as how users expect to integrate GenUI with their existing tools and practices.

**GenUI for All #1: Supporting Early-Stage Ideation**

GenUI can provide users with ideas or support them to come up with their own ideas in the early stage.

Participants of all roles considered GenUI a source of ideas for them to explore at the early stage of design, which can be particularly useful in highly-iterative design activities such as brainstorming and sprints (UXR1, UXD4, UXD7, SWE10). UXD2 would intentionally try different prompts just to lay out as many design possibilities as possible. Being able to generate UI screens at the early stage can also solve the "cold start" problem where the user has not familiarized themselves with the concept behind the design; GenUI can help to *"get something out and talking"* (UXD10):

> *As I didn't understand what mindful micro-activity was referred to, I kind of hoped the tool could just generate something for me to get started.* (UXR7)

Rather than passively receiving ideas from GenUI, participants used GenUI as a point of departure to come up with feature ideas or use cases that went beyond the ones generated by AI.

> *It also gives me some other ideas (even though it's not designed in the prototype), i.e, linking to external plant articles, linking to my own calendar or tasks app, etc.* (UXR3)

> *While looking at the initial design, [I] thought of a new app idea that I probably wouldn't have if I was just working on the PRD.* (PM3)

> *... was playing with different features and started to think of other use cases, e.g., what the screen should look like if the user has not logged in for a long time, which leads to many AI-generated ideas to enable this experience.* (UXD7)

These findings suggest that participants got value out of using GenUI at the early stage, which helped them get started and to continue moving through their design exercise.

**GenUI for All #2: Saving Time & Effort to Get the First Draft**

GenUI can save time and effort compared to current approaches of creating the "first draft" prototypes.

Participants acknowledged how much time GenUI could save them from manually creating a "first draft" UI prototype, improving productivity (PM6), and *"significantly reducing the time to generate baseline UX ideas"* (UXD6).

> *[GenUI] provide[s] the prototype for a whole app front-end for me, which previously I have been spend[ing] time with CSS/HTML styling [for] a long time ... after drawing the prototype in Miro.* (SWE12)

Participants also mentioned the value of GenUI for users working on a tight deadline where getting *"something good enough"* (SWE10) is the main goal and *"even if the generated interactions are imperfect, having a starting point with something already set up would cut the time needed"* (UXD3).

By automating the UI creation process, GenUI saves time for users to work on other aspects, such as usability and aesthetics (SWE12), which the user might have been too preoccupied to think of if they have to focus their efforts on using manual prototyping tools like Figma. By supporting the prototyping task conventionally done by UXD, GenUI also saves time for and thus accelerates other tasks: *"PM and UXD can spend more time on other stuff as opposed to spending too much time doing UI mocks"* (SWE8). In addition, PM7 and UXR1 pointed out how GenUI can shorten cycles (*e.g.*, visioning exercises) that require UXD's participation when there is limited UXD resources.

Besides quickly getting users to a first draft, many participants also recognized GenUI's time-saving benefits for automating routine and consuming design tasks, such as creating "standardized UIs" (UXD5) like a check-out screen or generating assets and contents that otherwise would take time to manually create or find (UXD5, SWE3). Although not yet available in the GenUI tool we tested, participants mentioned accessibility as a possible time-saving task GenUI can support, such as generating alt-text (UXD7) or checking adherence to accessibility guidelines (UXD5, UXD10). However, depending on the users' expertise in other design tools, GenUI might not save time for certain prototyping subtasks: UXD4 considered GenUI's ability to create UI from sketches not as fast as her crafting the same UI manually using Figma.

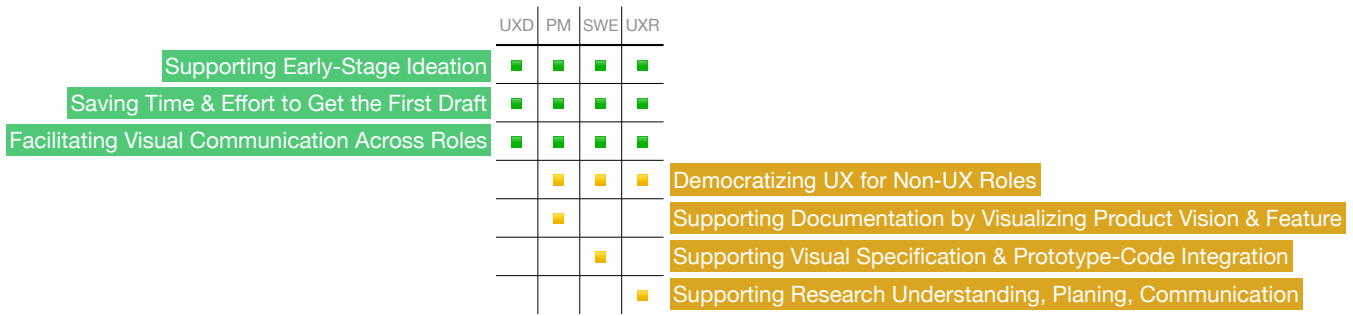

Figure 4: Our findings lay out promises and opportunities for GenUI to benefit all roles as well as providing specific support for individual non-UX roles.

> **GenUI for All #3: Facilitating Visual Communication Across Roles**
>
> GenUI can serve as a communication medium by providing visual aids to assist idea presentations across different roles and stakeholders

Participants across all roles valued how GenUI can help them visualize their ideas, *e.g.*, to "*visually style and format certain interactions*" (UXD7), to "*visualize the experience*" (PM2), and to "*visualize my vision for the app*" (UXR1).

> *Setting up the introduction flow was a great starting point. It helped me visualize potential features and the app's direction.* (UXD10)

> *[The generated] layout makes it easy to see how things connect to each other and gave a really quick visual of the kinds of apps I want the users to go through, which leads to requirements.* (PM7)

GenUI's visualization of design ideas, in turn, can facilitate communication across roles, getting ideas across, and "*align how different people think to be on the same page*" (UXD7). For example, working with a lot of data to make feature recommendations based on user behaviors, UXR5 mentioned that using GenUI can help visualizing the recommended features when presented to the team.

One participant's work involved complex dashboard that was often difficult to communicate to users; using GenUI can help "*convey our thoughts to the user more effectively*" (PM6).

> *... it gave me something tangible to work on—meaning showing the prototype to my colleagues to get some initial reactions to the features of the app.* (UXR7)

Current communication amongst stakeholders of different roles is limited to docs and slides that are "*too linear*" (UXR4). As a communication medium, GenUI can support roles that typically do not or cannot communicate via UI prototypes (UXD10), which encourages more participation in cross-functional planning and ideation workshops (UXR5).

> *When talking to designers I will come up with some ideas but am not good at drawing. [It] would be great to have someone (GenUI) draw the ideas for me.* (SWE8)

> *Some cross-functional partners want UXD to explore ideas but have no easy way to express what they have in mind; they can use this tool to communicate in ways that a designer would think.* (UXD3)

There were also suggestions for GenUI tool to eventually serve as the nexus of communication amongst multiple roles and stakeholders:

> *GenUI can keep records of all rounds of reviews/feedback, help connect all the feedback to the commenters and synthesize and give a list of action items.* (UXD2)

> *GenUI can allow for mak[ing] refinements and get feedback from users to finalize the look and feel and convey[ing] it to the software engineers via a shared canvas.* (PM6)

## 4.3 GenUI's Promises and Opportunities to Support *Non-UXD Roles*

While GenUI's original purpose might be to support UXD's work, we found that there are promises and opportunities for GenUI to target at non-UXD roles (Figure 4). Foremost, GenUI democratizes UX, making prototyping easier and more accessible to non-UXD roles; further, GenUI can tailor its support for specific non-UXD roles: PM, SWE, and UXR.

> **GenUI for Non-UXD #1: Democratizing UX for Non-UXD Roles**
>
> GenUI makes UX design more accessible to the broader non-UXD roles, allowing them to gain independence from relying on UXDs and accelerate their work.

Non-UXD roles sometimes create prototypes as part of their work, often in a fairly coarse, manual, and ad hoc approach, *e.g.*, copying and pasting screenshots of similar existing apps into slides (SWE9, PM6).

One shared perspective amongst participants is that GenUI lowers the bar for prototyping and was seen as a way to democratize UX for non-UXD roles, *i.e.*, people without formal training in design. Foremost, GenUI can provide lots of examples (UXD8) to fill in the UX knowledge gap (SWE7), such as showing someone unfamiliar with UX design what a typical check-out screen might look like (UXD5). As such, GenUI can make the creation of UI mock-ups "*a lot easier across every role*" (SWE8), thus allowing users to gain independence in performing low-stakes UX-related work. UXR7 felt that, in this study, she was doing what a UXD would typically do and another participant commented, "*I didn't have to rely on a UXD partner to generate the required screens for the study; I was able to do it myself*" (UXR1).

Multiple participants mentioned that gaining independence is important especially when there is limited UXD resource or it takes too long of a turn-around time to meet a deadline. With GenUI, teams can create UIs on their own to get approval. Some participants (UXR3, SWE7) even felt they could use GenUI and take on solo projects that otherwise would have to involve PM or UXD.

Gaining independence also allows non-UXD users to explore and pursue more ideas—

> *GenUI can help when I had a lot [of] design ideas that needed to be visualize[d]—turning thoughts into specific designs—but only have one designer who isn't available to work with me to realize all those ideas.* (UXR3)

One participant commented that GenUI tool should primarily seek to support non-UXD roles: "*Supply the team with design ideas; designer might not be the target audience. Leadership/engineers first; designers last*" (UXD1).

Although GenUI holds the promise to democratize UX, multiple participants did acknowledge that GenUI as-is still could not replace designers (UXR1, UXD7, UXD12) and a potential pitfall could be non-UXD overrelying on GenUI's imperfect output (detailed in §4.4), thus risking the design quality of the overall project. Further, the design of GenUI tool needs to keep in mind that non-UXD users might struggle either in learning all the UX-related features (UXR1), *e.g.*, editing options, or in composing a prompt using a UX-like language (SWE4).

> **GenUI for Non-UXD #2: Supporting PM's Documentation by Visualizing Product Vision & Features**
>
> GenUI can visualize product vision as well as specific feature details for PM to incorporate into a product requirement document (PRD).

A recurring GenUI capability mentioned by PMs (more so than any other role) is visualization. As PM7 pointed out, when creating a PRD, a PM needs to imagine what we can see at the end, which is usually unavailable in visual forms until a UX person creates the prototype later—GenUI can shorten this gap. It is generally not required (nor practical) for PRD to have visuals of product features; with GenUI, some PMs would like to include generated UIs in their PRD (PM1, PM5).

For PMs, one key benefit of GenUI's visualization of product features is reducing ambiguity to enable clearer requirement descriptions.

> *It was also helpful to see how words I used in the prompts translated into visuals, e.g., "modern" resulted in a very clean looking app, which is different from what I had in mind, so that's not a term that I should use when describing the app in PRD or to the team.* (PM3)

PM4 went further and considered it important to generate an animated prototype beyond static visuals:

> *Having an animated prototype can help in the feature definition phase because I can explicitly layout what exactly it looks like and work. In PRD, even with mocks there are lots of room of interpretation. Having an animated storyboard can reduce ambiguity and help stakeholders understand and interpret the PRD better* (PM4)

PMs also suggested having a more PRD-oriented GenUI output format, such as adding the generation of textual descriptions of the UI (PM7) or even presenting a "*requirements-based, annotated generated UI.*" (PM1) to replace PRD. This suggestion is related to expanding the modality of GenUI output, which we detail in §4.4.

> **GenUI for Non-UXD #3: Supporting SWE with Visual Specification & Prototype-Code Integration**
>
> GenUI can support SWE's work by providing visual specifications to guide coding, converting design to generated code, and integrating itself into the developing environment.

One perceived benefit for SWE is how GenUI can provide visual specifications as a framework to guide their coding work. Such visual support does not have to be of high-fidelity, as pointed out by one participant—it could simply be "*some square boxes as an abstract representation of the UI, which are enough for coding to start*" (SWE10). Given limited designer resources, SWE can quickly see what an idea would look like as opposed to building it from scratch (SWE6). Since committing to coding an idea could be risky, GenUI allows SWE to "*quickly bootstrap some ideas to discuss with people before actually coding*" (SWE11), which is suitable for small projects without designers (SWE4).

In order to realize support for SWE, future GenUI tools should enable better prototype-code integration, such as integrating GenUI features into IDEs (SWE9), pointing to changes made in code when modifying UI (SWE9), supporting code generation for various UI frameworks (SWE5, SWE6, SWE12), easy transfer of generated code (*e.g.*, CSS) to the development codebase (SWE5, SWE9).

> **GenUI for Non-UXD #4: Supporting UXR to Understand, Plan, and Communicate Research**
>
> GenUI can serve as visual aids to help UXR better understand product features, come up with research questions and plans, and communicate their research with co-workers.

Foremost, UXR's job requires a solid understanding of users and the corresponding product features. Since the task of our study is specific to researching features of a new app (rather than improving features of an existing app), UXR participants often found GenUI most useful to generate designs for them to understand how the UI might be like (UXR3), based on which they can start formulating research questions (UXR4). GenUI, as a source of UX knowledge, can inform UXR of best practices in industry when UXR has limited concrete ideas of current approaches (UXR7). As one participant summarized—

> *[GenUI] tool can help understanding the user flow and experience, major interaction features, and highlight potential friction points and pain points.* (UXR3)

As a next step, GenUI can help UXR create their study plan, such as creating a persona (UXR2), generating different text labels to test UX writing (UXR3), or including a generated storyboard with text description to present multiple options to gather users' reactions (UXR3). It would be useful to incorporate the kinds of research in the prompt, which should result in the integrated generation of a study plan with embedded UIs or UI screens embedded with dialogues of study questions (UXR6). Compared to their existing approach, a GenUI-assisted research plan "*can be very precise because you know the designs and what you want to test*" (UXR2).

Finally, as some UXRs are involved in other parts of the project cycle, GenUI can support them to communicate otherwise abstract research questions to stakeholders via the generated mock-ups (UXR2)—

> *Text-to-UI generation can work with UXR as a collaborator. Often to propose an idea, UXR needs to explain an idea, a feature, a product offering; UXR can go to the tool and generate something that matches what they have in mind and include it in the proposal. This is useful when UXR is involved in strategy development.* (UXR7)

### 4.4 Gaps in GenUI—Ongoing Challenges and Areas for Improvement

We identify the following gaps in GenUI's ability to realize the promises and opportunities of supporting UX practitioners and beyond. As shown in Figure 5, these gaps span the end-to-end process, from before using GenUI, to specifying input to GenUI, to issues related to GenUI's output, and to users' interacting with GenUI output. Although we distilled the following findings based on participants' experiences with one specific GenUI tool, we believe these gaps likely exist in other GenUI tools as well, given these tools' similar technical back-ends and interactive front-ends.

> **GenUI Gap #1: Problem Formulation with Context**
>
> GenUI should support users to formulate a design problem with broader product context before exploring design solutions.

Formulating a design problem is as important as coming up with the design solution. Existing GenUI tools, unfortunately, did not have targeted features to support problem formulation and some participants resorted to using language models before turning to the GenUI tool.

> *I first started with using Gemini to study which mindful micro-activity to focus on. This conversation was needed for me to understand more about mindful micro-activities, and to brainstorm which type of app I should create.* (UXD4)

Some participants did find the exercise of composing prompts got them to think about what they wanted. GenUI tools can further exploit the prompting step as a way to help users with problem formulation.

> *Writing out a prompt for what I wanted and choosing styles got me started thinking about what I wanted.* (SWE10)

> *Because I had to form a prompt to generate the customization screen, it helped me to think through feature requirements immediately (e.g., other than the ability to customize duration of break, X Y Z etc should be customizable as well).* (PM3)

> *The prompts forced me to think about what requirements and style I wanted more carefully.* (PM5)

Other participants considered a thought process should precede prompting—a step "*not needing a tool*" (PM3)—or instead having a "pre-prompt" feature, *e.g.*, an "*idea board*" (UXR4). By the end of this step, there should be a well-defined list of requirements before using GenUI (SWE10).

To some participants, problem formulation should also result in the broader product context beyond a sequence of UI screens. Currently, participants felt that GenUI would not be able to know all the specifics about the products they work on (SWE4) and the user needs to provide more project-level context to fully convey one's design intent (SWE9, PM2). As part of the design problem, it is helpful for GenUI to incorporate some historical context, *e.g.*, the team's previous design decisions (UXD1) by accessing team members' conversations and exchanged documents.

> **GenUI Gap #2: Assimilating Users' Intents**
>
> GenUI should assist users in expressing their intents—via or beyond prompting—while following explicit instructions and inferring implicit intents.

Participants noted that GenUI sometimes struggled to follow even explicit instructions, such as style keywords in the prompt

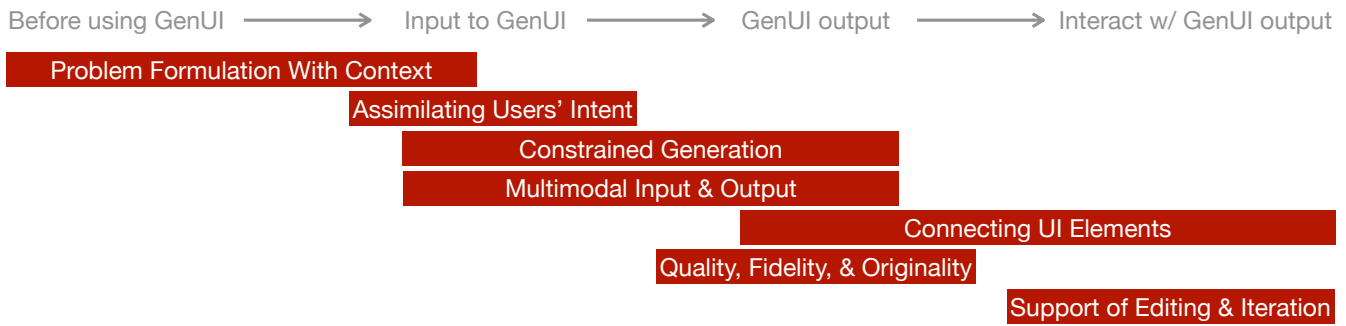

Figure 5: Our findings identified gaps—GenUI's ongoing challenges and areas for improvement–throughout the entire process of a user interacting with a GenUI tool.

(UXR3, UXD5, PM5) or uploaded screenshots as examples (UXD8). SWE10 gave a very detailed prompt but GenUI did not realize all those details and PM7 noticed extra generated elements irrelevant to his prompt. One participant summed up her prompting experience as GenUI being "*useful when not knowing what I want; but not so much when knowing exactly what I want*" (UXR7).

At times, a user could feel stuck in a prototyping step as GenUI was unable to perform the instructed UI generation despite users' multiple prompting attempts (UXD8, SWE12)—as one participant lamented, "*at some points it just stopped making changes regardless of my input*" (UXR3).

Participants also noticed that their prompt did not include all their design intents, some of which often remained implicit, making it harder for GenUI to interpret and realize. Participants mentioned how GenUI's output "*started to show some of my idea but not quite*" (PM3), did not capture what they "*had in mind*" (UXR2, UXR3 PM3) or match "*much of my product idea*" (UXD12), their high-level design goals (UXR2), or their mental model of what the app should be like (SWE1, SWE8).

> *Maybe AI didn't interpret the prompt fully. [I] might have eluded to it but might not have specifically described it.* (PM3)

> *The AI did not understand the prompt very well. It did understand parts of it like adding the "skip" and "next" [buttons], but not the core essence of the prompt.* (UXD6)

> *AI just pretended to understand me and modified the image without actually modifying it.* (UXD12)

These shortcomings were not necessarily GenUI's "fault". It seemed that prompting, like any other use of natural language, would tend to omit certain implicit intents—a common issue in other generative AI as well [3]. Some participants found prompting worked better when they intentionally tried to be more detailed (PM2); on the other hand, writing detailed prompts takes effort and participants felt they should not have to fully describe all their implicit intents, which "*defeats the purpose [of using GenUI]*" (UXD2).

Indeed, participants realized the challenges of composing the right prompts to express what they wanted GenUI to design. Some mentioned they made use of the existing GenUI tool features for composing the initial prompt, e.g., by inserting suggested style keywords or referring to examples of supported prompts (UXD5). However, help seemed more needed when they tried to perform specific subsequent actions, e.g., positioning an element (PM4).

Participants wished there were some assistance rather than trying different prompts on their own with no avail— "*I think if the way I worded it wasn't good then I could benefit from guidance on how to better write the prompt*" (UXR5). Instead of describing everything in an open-ended vocabulary, some suggested an alternative mode of choosing from a closed set of design options (UXR3), such as how many screens they want to create (UXD6). GenUI could also take more initiatives in asking for information needed (SWE3)— "*It could have asked me more questions about a persona, features, things I didn't want, etc*" (UXR3); or even start a conversation— "*The tool could generate something, ask what [the user] doesn't like about it to get smart enough to make the change requested*" (UXD1).

> **GenUI Gap #3: Constrained Generation**
>
> Rather than open-endedly, GenUI can constrain itself with pre-defined, specific modes of generation at the cost of being too limited; the generated UIs should follow constraints of organizations' design systems, specific application domains, and best practices for accessibility.

In terms of constraining input to GenUI, participants commented on the trade-offs between two pre-defined modes provided in the tool we tested: more exploratory *vs.* more precise. Some expected interacting with GenUI to be like talking to a chatbot rather than being limited to a few options (SWE5), which seemed "*too much like a formula*" (SWE12). Others, however, preferred a high-precision mode because it made prompting feel simpler (SWE12) and because they knew what they wanted (SWE7) and would like to have pre-defined prompts for specific design tasks (UXR3). Overall, there seemed to be winner across the two modes and GenUI should offer users options to invoke constrained vs. "*free-style*" (SWE12) generation on a case-by-case basis.

> *An open-ended input box would be very flexible but would require expertise to know what you are doing in*

*the first place; the limited options is great for beginners given specific steps.* (SWE5)

In terms of constraining GenUI output, participants were mostly concerned about GenUI's adherence to the company's design systems, *e.g.*, UI libraries with pre-defined styles, which could be the major roadblock for users to adopt GenUI results beyond just concepts (UXD8, UXR2). In order to use GenUI for their work, the tool should not create new components outside of the existing design system (UXD2) and should warn the user when their edits of UI do not meet the design system's requirements (PM3). Alternatively, it is also possible to provide an additional step that modifies generated UI elements based on design systems' requirements (UXD7).

Some participants also would like the ability to constrain the generated UI in the specific application domains they work on. Some of these participants' work was on internal tools for engineers whereas GenUI was mostly about consumer apps and might not be able to support their work (UXR1, UXD8). To bridge this gap, PM6 suggested cataloging more variety of use cases and UXD8 suggested a feature similar to in-context learning—a user would show GenUI how an existing screen addresses domain-specific requirements and then ask GenUI to come up with alternative designs.

Last but not least, another frequently-mentioned constraint for GenUI output was accessibility. Participants noticed accessibility issues in the generated UI elements that should have been automatically corrected (UXD1, UXD8). In addition, there could be more built-in support for adhering to best practices in accessibility, *e.g.*, adding a tooltip when hovering a button (SWE5).

> **GenUI Gap #4: Multimodal Input & Output**
>
> Rather than conforming to limited formats, users want to versatilely provide multimodal inputs to GenUI, which, in turn, should generate multimodal outputs for diverse needs.

In terms of GenUI input, beyond a one-size-fits-all text prompt, the tool's input modality should tailor itself to each role's ability. For example, using text prompts seemed a reasonable choice for PM whose work already involves using language to describe and communicate products (PM5). Sketch might work well for UXD who might otherwise spend more time expressing their ideas in a text prompt (UXD4) or for non-UXD roles who already used hand-drawn UIs to talk to designers (PM2, SWE5). A common request was having requirements as a specific input format (PM1) and some even expected GenUI tools to accept a much larger scale of requirement information as input, even an entire document like a PRD (PM5, UXD2). Another viable option—which already existed in many GenUI tools—is using existing examples (*e.g.*, screenshots) as input, which would allow a quick reference to certain styles (SWE1, PM6); UXD12 further pointed out that such "examples" could also come from an image generation AI. There were further suggestions besides the conventional text and image input. For instance, UXD12, based on how she ideated in her work, suggested providing a user flow diagram as input as well as annotations on generated UIs as subsequent input for GenUI to refine its output. UXD4 suggested video as another possible input format, in which the user can walk through a sequence of existing UIs to describe their style.

Given all these possible input formats, participants also hoped GenUI could allow for flexibly switching between or combining multiple modalities of input (UXD12, SWE12)—the most common use case is uploading a screenshot with a text prompt as seen in many chat-based language model interfaces. Further, GenUI tools should intelligently nudge the user to give examples or change modality from text to image, or to a reference screen from another app (UXR2).

On the output side, participants suggested numerous ways of expanding the scope of current GenUI results beyond a single modality of visual UI screens. One common theme was generating a textual version of the UI screens, including a one-paged product summary to convince stakeholders or as a hand-off to designers (PM5). It is also useful to generate a list of bullet points of to summarize how a user would interact with a screen, which helps creating a critical user journey (PM7). As mentioned earlier in GenUI's support for UXR, an alternative output modality would be a template-based study plan that includes an introduction and the generated UI screens (UXR1, UXR2, UXR6). Others would like the GenUI tool to provide technical information as part of its output, including different browser specifications to help cover edge cases (SWE7), warnings of violating any company design standards (UXD4), and accessibility-related labels, *e.g.*, alt-texts (SWE5).

> **GenUI Gap #5: Connecting UI Elements**
>
> GenUI should meaningfully connect otherwise isolated UI elements and screens with consistent styles, shared application contexts, proper hierarchical layouts, and coherent user flows.

Accordingly to participants, generated UI elements were seldom independent of each other; rather, GenUI tool should connect UI elements within or across screens. Foremost, connecting UI elements across screens means applying consistent styles (UXD5, UXD6, PM1) and those generated later should refer to the earlier ones in styling (UXD1, UXD3, SWE11) as well as fitting in the existing layout (UXD4, SWE9) and user flow (UXR5). In addition, GenUI tools should allow users to apply a style on UI elements across all screens (SWE1). Unfortunately, the GenUI tool we tested currently seemed to struggle with maintaining such "connective tissues"—

> *I had specifically asked it to reference my current chosen design, which it did not seem to do.* (UXD6)

> *I noticed that when adding new screens, it did not keep the basic tab structure design the same. It feels like it doesn't keep the design system consistent across newer screens being added.* (UXD7)

Perhaps one reason of inconsistency is a lack of shared contextual information that would connect generated UI elements under unified styles and structures.

> *I found that the AI assistant is not very contextual, and it will only give responses based on the exact prompt given, and not take into consideration the page that* *we're working on. Prompts therefore need to be very specific.* (SWE9)

As a result, some participants felt the need to re-elaborate contextual information in subsequent prompts, which felt repetitive and increased effort. For example, some participants had to always specify designing for a mobile app in their prompts.

> *Every time I had to add a screen i had to define the param[eter]s of it, but it was already established what kind of app it was from the beginning (mobile, etc.)* (SWE1)

> *When I had already selected ... that I want to design mobile app, it should not be asking that again ... while giving the prompt. It is kind of implied.* (UXD6)

Connecting UI elements also means appropriately lay out, organize, and update these elements based on their relationships with each other, such as grouping similar elements (SWE3, UXD10), possibly in a hierarchy (UXD1, UXD12), and extracting their common style information (rather than as repetitive individuals) (SWE5). When users make a change of a generated UI element (*e.g.*, the displayed name of a button), the tool should make necessary updates across all screens that contain the same element (UXD3). GenUI tools might skip the important step of propagating updates through UI elements supposed to connect with each other. For example, UXD2 asked to remove some generated elements in a 3×3 grid and the GenUI tool did remove those elements but without updating the grid layout, leaving awkwardly-looking "holes" amongst the remaining elements.

Finally, connecting UI elements also means putting them together into a coherent user flow. From a user flow perspective, some participants at times noticed missing screens in the generated sequence of UIs (PM3) and would prompt GenUI to add more screens to explore or change the flow (SWE1). Participants emphasized the need for GenUI tools to support generating user flow beyond creating a collection of disconnected UI screens and elements.

> *Need a hero user story using this app in a coherent flow. Just the screens only is too technical. Need to organize the generated UI screens in a way as a user is interacting with it.* (UXD3)

Participants expected GenUI tool to let them connect not just from one screen to another but also from one specific element to another (UXD7) or even to automatically suggest such connection based on an understanding of UI elements' semantics (UXD3) or a user-provided user flow diagram (UXD12).

User flow support is also useful for non-UXD roles, *e.g.*, for UXR to discuss with users about their journey in using the app (UXR2). However, unlike UXD familiar with other design tools for creating user flow (*e.g.*, Figma), non-UXD roles would need more support from GenUI tools (UXR7).

> *I am not an interaction designer, so I didn't find it easy to map the interaction flow. This could also potentially be a generative feature - I explain how the flow should look like and it automatically adds the interaction flows.* (UXR5)

> **GenUI Gap #6: Quality, Fidelity, & Originality**
>
> At present, Generated UIs need to improve their quality, provide an appropriate fidelity, and support original problem solving.

When assessing GenUI output, participants' feedback spanned three key aspects: quality, fidelity, and originality.

Participants often noticed quality issues in the generated UIs, such as truncated text (UXR3) in a button or misalignment of UI elements (UXD6). Although some of these imperfections seemed small (and common in other generative AI models), they might distract end-users when testing the generated UI screens (UXR3) and fixing them would cause interruption and add friction to the experience of using GenUI tools (UXD10). Participants acknowledged that, at present, GenUI quality was not production-ready (UXD7) and would require further tweaking and polishing (UXD3). We discuss later (§5) about the right expectation of GenUI quality to utilize its capability appropriately.

Relatedly, participants commented on the right fidelity of the generated UIs. Although the default GenUI output is a high-fidelity prototype, UXDs pointed out that low fidelity was more appropriate for early stages of design (UXD1, UXD2, UXD8). Some non-UXD participants also considered a lower fidelity appropriate for their work. For example, one participant preferred a wireframe because they can "*edit it quickly without much learning curve*" (PM5) and UXRs considered a low-fidelity screen enough for study planning (UXR1) and for eliciting users' reactions to different features (UXR3). Even for the same role, the preferred fidelity could differ. For example, some SWE would prefer a high-fidelity output rather than wireframes (SWE1) whereas others might consider wireframes good enough to get coding started (SWE10).

Notably, participants brought up a critical point about how high-fidelity generation might be counterproductive to design idea exploration.

> *The high-fi generation pins me down with too much "doneness" with a tendency to just use it as-is. It's hard to say to get rid of that and restart, showing me things too close to done too soon, influencing [my] ability to branch out and explore.* (SWE10)

> *Generated UI can provide the visuals but might limit participants' imagination; text could have the power for participants to imagine what the UI might be like.* (UXR3)

One deeper concern from participants was the originality of GenUI output. An intrinsic challenge for GenUI is the ability to accurately understand the design problem statement (UXR6) and to thoughtfully solve the underlying problem, *e.g.*, achieving the goal of getting more users to try a product (UXD2). As one participant put it, GenUI needs to "*[provide] more creative or thoughtful addition: think about the problem the sketch represents and design based on such understanding of the problem*" (PM2). Some noted GenUI seldom came up with truly original solutions beyond generating seemingly templated UIs that lacked specificity to users' product in mind— "*[The generated UIs] could fit any other app ...*" (SWE3).

Although generated UIs as-is might not provide a high-quality or original solution, some participants were able to see past such limitations and build on the imperfect results either as a "*brainstorming partner*" (PM1) to focus on exploring the problem space or to think more and improve the given solution, *e.g.*, "*being able to poke around*" (PM4) to cover some edge cases or to detect issues of the initial design (PM5). To these participants, GenUI's output is simply an intermediate step to build their own thinking on.

> *Seeing the initial visuals helped me to think more deeply about my idea and to refine the vision.* (PM3)

> **GenUI Gap #7: Support for Editing & Iteration**
>
> Users need support for editing the generated UI elements within and across screens, thus enabling rapid, low-cost, and informed iterations to explore different designs.

One of the most common participant feedback was that current GenUI tool's lack of support for editing individual generated UI elements, which in turn added friction to the iterative nature of the design process (PM1). The difficulty of editing is exacerbated by the amount of editing users often felt the generated UIs needed and, as a result, they would give up editing and instead restart generating a new sequence of UI screens.

> *It takes too much to change since it's quite different from what I wanted at the beginning. I ended up thinking that I should start over.* (SWE8)

> *[I] didn't find a way to refine the initial screens generated [and] need to start from scratch instead.* (PM6)

Participants also suggested interaction design ideas for GenUI tools to make editing easier, such as a "find and replace" feature to fix a common issue across multiple UI elements (PM1), and easier placements by snapping generated UI elements on a grid made up of 10*px* cells (UXD6).

Besides individual UI elements, participants also mentioned the challenge of prompting GenUI to support editing an entire screen or elements across multiple screens, such as enabling a single action for changing the theme or style applied to all screens rather than using prompts for each individual one (UXD2, SWE7, SWE8). Some participants needed the GenUI tool to blend multiple screens (PM4) or, conversely, to split a busy screen into multiple ones (SWE1). Another set of requests for GenUI was the ability to select (SWE3) or move (UXD4) UI elements across screens.

All these above editing activities are common in direct manipulation based design tools and participants expected them to be an integral part of a GenUI tool. Sometimes, when the changes are small (SWE12) or very specific (UXD10), it might be easier to support users' manual direct manipulation than prompting AI to perform the same actions (UXD6, SWE1, PM4, UXD12).

> *I wanted to copy the menu bar design from another page to this. I had to do the selection plus the copy-pasting myself. The [AI] also did not understand me trying to create an identical menu bar at the top.* (SWE3)

> *[I] spent a lot of time trying to place the camera icon next to the picture. I want to put it next to the avatar but instead on top. I only realized I can manually re-position it.* (PM4)

Because one of GenUI's promises is to democratize UX, its support for the necessary UI element editing also needs to align with this mission. The challenge is how to make editing even easier for non-UXD users who might not be familiar to direct manipulation based design tools (UXR4, SWE10).

Support for editing begets the ability to facilitate rapid, effortless iteration—changing things quickly to see the effects (SWE6), which allows users to compare different design options and parameters (UXD3, UXD7) or their rationale (SWE5). Some participants related to how they would interact with chat-based language or image-generation models—iteratively editing the same prompt to see how the results differ or improve (SWE4).

> *In Midjourney, I can easily modify my prompts and get new images in the same feed. I would give multiple prompts to Midjourney and do something else while waiting for the results of all the prompts.* (UXD12)

The GenUI tool we tested had a latency that participants felt too long, compared to how they could "*flip through maybe 20 [existing] designs to get inspiration rather than wait[ing for] one [generated] design*" (SWE10). There was only one sequence of UI screens generated at a time, which could make participants feel "*stuck*" (SWE10) or want to "*quickly restart*" (SWE8) as soon as seeing the result not like what they had in mind.

Participants also hoped GenUI tool could visualize how design iterations evolve over time for comparing different designs of the same feature (SWE8) or to show how designs address insights from other activities (*e.g.*, user research) in-between iterations (UXR4). Some would like to revisit or revise earlier prompts to create a new branch of iterations (PM3, SWE12).

## 5 Discussion & Implications for Design

In this section, we distill key insights from our findings and discuss their implications for the design of GenUI tools.

### 5.1 The GenUI Verdict: Good First Draft, Tough Last Mile

At present, the most salient benefit of GenUI is helping users quickly and effortlessly get a good (enough) first draft, overcoming the inertia of getting started at the early stage of a project. However, the generated UI elements often suffer from various degrees of quality issues[16], which creates a "last mile" problem, requiring a non-trivial amount of editing work to go from a GenUI first draft to a production-ready version of design.

There are two directions to address this "last mile" problem. The first one is focusing on improving the back-end models. Since many GenUI models build on code generation AI (*e.g.*, [11]), which has been showing continuous improvement, we can reasonably expect a higher quality of generated codes that render UI elements in the future. The other approach is to provide better tool support for editing and iteration of the first-draft UI, as requested by many of our participants. However, it is questionable whether such editing and iteration should take place in a GenUI tool or in other tools that already exist precisely for such purposes (*e.g.*, Figma). Rather than adding more user interaction, perhaps GenUI tools should stay lightweight, focus on what they do best (providing first drafts), and hand off further editing tasks to other tools. To prepare for such hand-off, GenUI tools can connect otherwise isolated generated UI elements to maintain consistent styles, proper groupings, and hierarchical layouts.

---

[16]We believe such issues are not unique to the tool we tested or to the GenUI family of AI; rather, quality is a common problem in most current generative AI models (most notably image generation AI that often creates incorrect artifacts, *e.g.*, unreal anatomy of human hands).

## 5.2 Not Just for Designers: GenUI Tools Can and Should Support Roles Beyond UX

GenUI might have been created to automate UXD's work, yet our findings suggest that it can and should support more roles beyond designers. Foremost, similar to how word processing software democratized writing and publishing, we believe that GenUI tools will democratize UX design for all. Therefore, the design of GenUI tool should foremost target at multiple roles, which means providing role-specific interaction modes and/or imbuing users' role-related information to adapt the model's generative behaviors.

To support broader roles, we believe it is important to expand GenUI's current input and output modalities. In terms of input, as various roles might "speak different languages," GenUI tools should support a wide range of input modalities beyond a singular form of text prompt, *e.g.*, sketch, video demonstration, requirement document, and user flow diagram. Likewise, in terms of output, the default format of high-fidelity prototypes is insufficient to meet different roles' needs. As prior work suggested, there should be different kinds of prototypes serving for varied purposes [10, 18]. GenUI should include other modalities, such as low-fidelity layout, study plan, user journey, and feature summary, and flexibly support users to pick and choose a custom output format.

## 5.3 The Untapped Potential: GenUI Tools Can Further Provide Team-Level Support

Prior studies [14, 15, 30] have pointed out the dual purposes of prototyping: exploration and communication. It seems that current GenUI tools primarily focus on supporting individual users' design exploration and there is untapped potential to enable team-level support, such as communicating designs with different stakeholders. As reported in our findings, GenUI tools can provide such support for communication by providing visuals to illustrate otherwise abstract text-only presentation or by annotating UI screens to better explain the visual-only user flow.

Further, future GenUI tools can even serve as the nexus—providing a shared workspace that connects, synchronizes, and integrates multiple team members' individual or group work. To start, proper tool designs can employ GenUI to facilitate collaborative work, *e.g.*, helping two users of different roles translate their ideas into generated UI as a shared visual medium. At a larger scale, GenUI can listen to and materialize a team's evolving discussions to specific UI updates, which requires further development that enables GenUI to extract broader product-related context and incorporate such information into the UI generation process.

## 5.4 Reality Check: GenUI Tools Should Address Practical Considerations

Our findings suggest that GenUI tools cannot succeed in a vacuum without addressing some practical considerations, which might not seem like major issues but will likely prevent users from adopting GenUI in their work. First of all, GenUI needs to find its place in a user's workflow and facilitate seamless transfer of work products to-and-fro their existing tools. The need of workflow integration echoes findings from prior work [20] that pointed out how "simplistic automation" by AI is unlikely to work due to a lack of integration into UX designers' existing workflow. Second, generating *any* UI has little practical use because teams and organizations often have their own design systems and GenUI tools need to adopt the standardized styles, layouts, and interaction flows. Similar to most language models, GenUI likely learns from Internet data of UI codes and needs to support organizations to fine-tune its models for specific design systems. Thirdly, also related to organization-specific needs, GenUI tools should expand its design knowledge beyond common consumer apps and support UI generation for specific application domains, such as dashboard for specialized data and internal diagnostic tools. Last but not least, GenUI tools should adhere to best practices for accessibility by adapting UI elements, adding accessible information, and warning designers of any existing accessibility issues.

## 6 Conclusion

GenUI is a new class of generative AI models that will play a key part in shaping the future of UX design in industry. Our research contributes empirical findings that, grounded in participants' hands-on experiences, explore the design of GenUI tools to support not just UX designers but other UX-related roles as well. We believe that GenUI's success lies in its unique ability to rapidly get users to a good first draft of design, democratizing UX for all while having the potential to provide role-specific support. To realize all these promises, however, GenUI does have to address a series of practical considerations, such as better integration with existing tools and organizations' design systems to close the "last mile" towards a product-ready design. The main limitation of the present study is the use of only one GenUI tool, which we plan to address by conducting follow-up studies to include other (newer) GenUI tools and reflect on this paper's findings based on new data. Another limitation is that the tasks only involved individual work; in the future, we plan to investigate how GenUI tools can facilitate collaboration among team members of different roles. To materialize this study's design implications, we are also working on developing new GenUI tools, which we we will evaluate by conducting a similar project-based study with various UX-related roles.

## Acknowledgments

We thank our collaborators and study participants for their contribution to this research. Savvas Petridis, Anna Kipnis, Mahima Pushkarna, Peggy Chi, and Michael Terry provided valuable feedback and suggestions to improve this work. We are also thankful for reviewers who provided valuable feedback to help us improve the paper.

## A Additional Information About Participants' Experiences with Generative AI

The following tables outline our study participants' prior experiences with generative AI, including language and image generation models.

There were 7 participants who reported prior experiences using GenUI-related tools, such as Vercel's v0, Claude Artifacts, Gemini, ChatGPT, Figma's "First Draft" feature, Uizard, Musho AI, Galileo AI, and Genius AI. Their use cases varied, 2 participants mentioned casual exploration rather than regular usage; 2 others highlighted active experimentation of GenUI in design workflows; 1 explicitly stated dissatisfaction with the output quality and another expressed a focus on staying informed about advancements in AI design tools as a professional necessity. These participants' GenUI experiences were mainly exploratory and none had used a GenUI tool for their work as an integral part of a project.

## B Codebook After the Initial Rounds of Analysis

We employed a grounded theory approach [1] to code the qualitative data (responses in both the daily journal and the final interview) through multiple iterations: *(i)* The first author broke the data into atomic segments and labeled each with low-level codes to identify the underlying concept(s) related to our research questions; *(ii)* The first author reviewed the initial codes and developed them into higher-level codes, which we cataloged and defined in a code book; *(iii)* The second author reviewed the code book and discussed with the first author to resolve observed ambiguities, conflicts, and disagreements. By this end of these iterations, the code book contained 28 high-level codes as follows.

**Table 1: Participants' prior experiences using language models (*e.g.,* ChatGPT and Gemini).**

|                    | UXD | UXR | PM | SWE |
|--------------------|-----|-----|----|-----|
| Daily              | 7   | 6   | 3  | 8   |
| A few times a week | 4   | 1   | 2  | 3   |
| A few times a month| 0   | 0   | 2  | 0   |
| Rarely or never    | 0   | 0   | 0  | 1   |

**Table 2: Participants' prior experiences using image-generation models (*e.g.,* DALL-E and Midjourney).**

|                    | UXD | UXR | PM | SWE |
|--------------------|-----|-----|----|-----|
| Daily              | 2   | 0   | 0  | 1   |
| A few times a week | 4   | 1   | 3  | 4   |
| A few times a month| 5   | 2   | 4  | 6   |
| Rarely or never    | 0   | 4   | 0  | 1   |

- `gen ui quality`: GenUI should focus on improving the quality of its outputs, addressing even minor issues to minimize friction and enhance user experience
- `the last mile problem`: Although GenUI provide a good-enough first draft, it seems to require significant effort to close the "last mile", i.e., editing them to be product- or engineering-ready.
- `the promise as a first draft`: GenUI tools can help users create quick and effective first drafts of their designs that they can refine and build upon.
- `communication`: GenUI can serve as a communication medium by providing visual aids to assist idea presentations amongst UX and non-UX people
- `early-stage ideation`: GenUI can provide users with ideas or support them to come up with ideas in the early stage
- `time saving`: Using GenUI tools can save users' time and effort compared to their current approaches of creating prototypes
- `visualization`: GenUI tools can help visualize users' design ideas or serve as visual aids for role-specific purposes as well as communication
- `democratizing ux`: GenUI makes UX design more accessible to the broader non-UX people, allowing them to gain independence from relying on UXDs and accelerate their work
- `support for dev`: GenUI can support Dev's work by providing low-fi prototypes to guide coding, converting design to generated code, and integrating itself into coding environment.
- `support for pm`: GenUI can visualize product vision as well as specific feature details for PM to incorporate into requirements documents.
- `support for uxr`: GenUI can serve as visual aids to help UXR better understand product features, come up with research questions and plans, and communicate their research with co-workers
- `gen ui workflow`: Users follow different workflows when using GenUI for different purposes related to prototyping
- `workflow integration`: GenUI tools should be integrable with how users currently work and what tools they currently use in their work.
- `ai vs. human`: GenUI cannot replace human designers' input, such as their understanding of the requirements and design ideas
- `explainability`: GenUI should offer clear explanations of its capabilities, limitations, and rationale behind generated outputs, especially when results are unexpected.
- `human agency`: Users should calibrate their trust of and rely on GenUI appropriately; otherwise, overreliance might lose one's original thinking and result in suboptimal work products.
- `human-ai co-creation`: People can work with GenUI like fellow human co-workers and divide the labor to take on specific parts in UI prototyping.
- `assimilating intent`: GenUI should assist users in expressing their intents (e.g., via prompting) while following explicit instructions and inferring implicit intents
- `constrained generation`: Rather than open-endedly, GenUI can constrain itself with pre-defined, specific modes of generation at the cost of being too limited; the generated UIs should follow constraints of accessibility and design systems.
- `context`: GenUI should tailor the generated UI elements to the app's broader context, including its entire screens and the product-related background.
- `multimodal input/output`: Users want to provide multimodal inputs to GenUI and needs to generate multimodal outputs
- `problem formulation & solving`: GenUI tools should support users to explore and formulate a design problem and to come up with original solutions to the problem
- `representation & organization`: GenUI elements should be modularized, appropriately grouped, and laid out hierarchically to support easy selection, position, and removal.
- `support iteration & exploration`: GenUI should support users' rapid, low-cost, and informed design iterations while exploring different design parameters and options.
- `support of editing`: GenUI should support users to efficiently edit generated UI or other contents at various granularity– both as individual elements (within or across screens) and as the entire screen.
- `ui fidelity`: The fidelity of generated UIs should be appropriate for the intended stage of support and the specific roles of the user
- `user flow`: GenUI tools need to support the creation of user flow that connects multiple screens into an animated, meaningful user story

- `usability`: Current GenUI tools have usability issues that add to learning difficulties, confusions, and frictions to the usages of GenUI.

## C Post-Study Semi-Structured Interview Protocol

**Objective:** To gather qualitative data on participant experiences with the GenUI tool post-study.

**Format:** One-on-one semi-structured video interviews.

**Duration:** Approximately 30 – 60 minutes, variable based on participant engagement.

**Consent:** Oral informed consent obtained at the start of each interview.

**Recording:** No audio/video recording. Interviewer performed manual transcription via detailed note-taking.

**Procedure:**
(1) **Work Product Review:** Participant briefly presented their work created using GenUI.
(2) **Daily Log Review:** Walkthrough of participant's daily Google Slide logs, with interviewer asking clarifying and follow-up questions.
(3) **Overall Feedback:** Discussion of perceived opportunities and concerns for GenUI regarding individual and collaborative workflows.

**Data Capture:** Interviewer manually transcribed notes directly into the participant's respective Google Slide file containing their daily logs.